\documentclass{article}
\usepackage{a4wide}
\usepackage{amsmath}
\usepackage{amssymb}

\newcommand{\refap}[1]{Appendix~{\ref{#1}}}

\newtheorem{theorem}{Theorem}
\newcommand{\reftheo}[1]{Theorem~{\ref{#1}}}

\newtheorem{lemma}{Lemma}
\newcommand{\reflemma}[1]{Lemma~{\ref{#1}}}

\newtheorem{fact}{Fact}
\newcommand{\reffact}[1]{Fact~{\ref{#1}}}
\newtheorem{definition}{Definition}
\newcommand{\refeq}[1]{(\ref{#1})}

\newcommand{\oracle}[1]{\mathcal{O}[#1]}
\newcommand{\super}[1]{\boldsymbol{#1}}
\newcommand{\affine}[1]{\super{\ball{#1}}}
\newcommand{\ball}[1]{\overline{#1}}

\newcommand{\half}{\mbox{$\frac{1}{2}$}}
\newcommand{\halfsqrt}{\frac{1}{\sqrt{2}}}

\newcommand{\C}{\mathbb{C}}
\newcommand{\R}{\mathbb{R}}
\newcommand{\N}{\mathbb{N}}

\newcommand{\Hi}{\C^{\{0,1\}}}
\newcommand{\PM}[2]{\mathrm{Pr}^{#1}[{#2}]}
\newcommand{\BB}{\mathcal{B}}
\newcommand{\BS}{\mathcal{S}}
\newcommand{\U}{\mathrm{U}}

\newcommand{\SO}{\mathrm{SO}}
\newcommand{\ox}{\otimes}
\newcommand{\ket}[1]{| #1 \rangle}
\newcommand{\bra}[1]{\langle #1 |}
\newcommand{\braket}[2]{\langle #1 | #2 \rangle}
\newcommand{\ketbra}[2]{| #1 \rangle\langle #2 |}
\newcommand{\norm}[1]{|\! | #1 |\! |}
\newcommand{\normone}[1]{\norm{#1}_{1}}
\newcommand{\normso}[1]{\norm{#1}_{\infty}}

\newcommand{\dist}{\mathrm{dist}}

\newcommand{\distso}{\dist_{\infty}}

\newcommand{\abs}[1]{| #1 |}

\newcommand{\Tr}{\mathrm{Tr}}
\newcommand{\proof}{\noindent{\bf Proof: }}
\newcommand{\qed}
{\mbox{}
\nolinebreak
\hfill
\rule{2mm}{2mm}
\medbreak
\par}
\newcommand{\eps}{\varepsilon}
\renewcommand{\phi}{\varphi}

\newcommand{\dens}[4]{\left({
\begin{array}{cc}
#1 & #2 \\
#3 & #4 \\
\end{array}
}\right)}

\renewcommand{\Re}{\mathrm{Re}}
\renewcommand{\Im}{\mathrm{Im}}

\renewcommand{\neg}{{\mathrm{NOT}}}
\newcommand{\cnot}{{\mathrm{c\text{-}NOT}}}
\newcommand{\etal}{\mbox{et al.}}

\newcommand{\positive}[1]{\{1,2,\ldots,#1\}}

\begin{document}
\title{\textbf{Self-Testing of Universal and Fault-Tolerant\\
              Sets of Quantum Gates}}

\date{}
\author{
\emph{Wim van Dam}\thanks{C.W.I. Amsterdam; Centre
for Quantum Computation, University of Oxford.
{\tt wimvdam@qubit.org}}
\and \emph{Fr\'ed\'eric Magniez}\thanks{Universit\'e Paris Sud, LRI.
{\tt magniez@lri.fr}}
\and \emph{Michele Mosca}\thanks{University of Waterloo;
Centre for Quantum Computation, University of Oxford.
{\tt mosca@qubit.org}}
\and \emph{Miklos Santha}\thanks{CNRS, Universit\'e Paris Sud, LRI.
{\tt santha@lri.fr}}
}

\maketitle
\begin{abstract}
We consider the design of self-testers for
quantum gates.
A self-tester for the gates $\super{F}_1,\ldots, \super{F}_m$
is a classical procedure that, given any
gates $\super{G}_1, \ldots, \super{G}_m$,
decides with high probability if each $\super{G}_i$ is close to
$\super{F}_i$.
This decision has to rely
only on
measuring in the computational
basis the effect of iterating the gates on the classical states.
It turns out that instead of individual
gates, we can only design procedures for families of gates.
To achieve our goal we borrow some elegant ideas of the
theory of program testing:
we characterize the gate families by
specific properties, we develop a theory of robustness for
them, and show that they lead to self-testers.
In particular we prove that the universal and fault-tolerant
set of gates consisting of a
Hadamard gate, a $\cnot$ gate, and a
phase rotation gate of angle
$\pi/4$ is self-testable.
\end{abstract}

\section{Introduction}
In the last decade quantum computing has become an extremely active
research area. The initial idea that the simulation of quantum
physical systems might be out of reach for classical devices goes
back to Feynman\cite{Fey82}. He raised the possibility that
computational devices based on the principles of quantum mechanics
might be more powerful than classical ones. This challenge to the
quantitative version of the Church-Turing thesis, which asserts that
all physically realisable computational devices can be simulated
with
only a polynomial overhead by a probabilistic Turing machine, is
the driving force behind the study of quantum computers and
algorithms.

The first formal models of quantum computing, the quantum Turing
machine and quantum circuits were defined by
Deutsch\cite{Deu85,Deu89}. Yao has shown\cite{Yao93} that these
two models have polynomially equivalent computational power when
the circuits are uniform.
In a sequence of papers oracles have been exhibited
\cite{DJ92,BB92,BV97,Sim97},
relative to which quantum Turing machines are
more powerful than classical (probabilistic or non-deterministic) ones.
These results culminated in the seminal paper
of Shor\cite{Sho97} where he gave polynomial time quantum
algorithms for the factoring and the discrete logarithm problems.

A quantum circuit operates on $n$ quantum bits (qubits), where $n$
is some integer. The actual computation takes place in the Hilbert
space $\C^{\{0,1\}^n}$ whose computational basis consists of the
$2^n$ orthonormal vectors $\ket{i}$ for $i \in {\{0,1\}}^n.$
According to the standard model, during the computation the state
of the system is a unit length linear superposition of the basis
states. The computational steps of the system are done by quantum
gates which perform unitary operations and are local in the sense
that they involve only a constant number of qubits. At the end of
the computation a measurement takes place on one of the qubits.
This is a probabilistic experiment whose outcome can be $0$ or $1$,
and the probability of measuring the bit $b$ is the squared length
of the projection of the superposition to the subspace spanned by
the basis states that are compatible with the outcome. As a result
of a measurement, the state of the system becomes this projected
state.

The most convenient way to describe all possible operations on a
quantum register is in the formalism of `density matrices'. In this
approach, which differs from the Dirac notation, the quantum
operations are described by completely positive superoperators
(CPSOs) that act on matrices. These density matrices describe mixed
states (that is, classical probability distributions over pure
quantum states), and the CPSOs correspond exactly to all the
physically allowed transformations on them. Such a model of quantum
circuits with mixed states was described by Aharonov, Kitaev and
Nisan\cite{AKN98}, and we will adopt it here. The unitary quantum
gates of the standard model and measurements are special CPSOs.
CPSOs can be simulated by unitary quantum gates on a larger number
of qubits, and in \cite{AKN98} it was shown that the computational
powers of the two models are polynomially equivalent.

Unitary quantum gates for small number of qubits have been extensively
studied. One reason is that although quantum gates for up to three
qubits have already been built, constructing gates for large
numbers seems to be elusive. Another reason is that universal sets
of gates can be built from them, which means that they can simulate
(approximately) any unitary transformation on an arbitrary number
of qubits. The first universal quantum gate which operates on three
qubits was identified by Deutsch\cite{Deu89}. After a long sequence
of work on universal quantum gates
\cite{DiV95,Bar95,DBE95,Llo95,BBC95,Sho96,KLZ96,Kit97}, Boykin et
al.\cite{bmprv99} have recently shown that the set consisting of a
Hadamard gate, a $\cnot$ gate, and a phase rotation gate of angle
$\pi/4$ is universal. In order to form a practical basis for
quantum computation, a universal set must also be able to  operate
in a noisy environment, and therefore has to be
fault-tolerant\cite{Sho96,AB97,Kit97,KLZ98}. The above set of three
gates has the additional advantage of also being fault-tolerant.

Experimental procedures for determining the properties of
quantum
``black boxes'' were given by Chuang and
Nielsen\cite{CN97} and
Poyatos, Cirac and Zoller\cite{PCZ97}, however these procedures
implicitly require apparatus that has already been tested and
characterized.

 The idea of self-testing in quantum devices is
implicit in the work of Adleman, Demarrais and Huang\cite{ADH97}.
They have developed a procedure by which a quantum Turing machine
is able to estimate its internal angle by its own means under the
hypothesis that the machine is unitary. In the context of quantum
cryptography Mayers and Yao\cite{MY98} have designed tests for
deciding if a photon source is perfect. These tests guarantee that
if source passes them then it is adequate for the security of the
Bennett-Brassard\cite{BB84} quantum key distribution protocol.

In this paper we develop the theory of self-testing of quantum gates
by classical procedures. Given a CPSO $\super{G}$ for $n$ qubits,
and a family $\mathcal{F}$ of unitary CPSOs, we would like to
decide if $\super{G}$ belongs to $\mathcal{F}$.
Intuitively, a self-tester is a procedure that answers the question
``$\super{G}\in \mathcal{F}$ ?'' by interacting with the CPSO $\super{G}$
in a purely classical way. More precisely, it will be a probabilistic
algorithm that is able to access $\super{G}$ as a black box in the
following sense: it can prepare the classical states $w\in\{0,1\}^n$,
iterate $\super{G}$ on these states, and afterwards, measure in the
computational basis.
The access must be seen as a whole, performed
by a specific, experimental oracle for $\super{G}$: once the basis
state ${w}$ and the number of iterations $k$ have been specified,
the program in one step gets back one of the possible probabilistic
outcomes of measuring the state of the system
after $\super{G}$ is iterated $k$-times on ${w}$. The intermediate
quantum states of this process cannot be used by the program, which
cannot perform any other quantum operations either. For $0
\leq \delta_1  \leq
\delta_2$, such an algorithm will be a $(\delta_1,
\delta_2)$-tester for $\mathcal{F}$ if for every CPSO $\super{G}$,
whenever the distance of $\super{G}$ and $\mathcal{F}$ is at most
$\delta_1$ (in some norm), it accepts with high probability, and
whenever the same distance is greater than $\delta_2$, it rejects
with high probability, where the probability is taken over the
measurements performed by the oracle and by the internal coin
tosses of the algorithm.
Finally we will say that $\mathcal{F}$ is
{\em testable} if for every $\delta_2 > 0$, there exists $0 <
\delta_1 \leq \delta_2$ such that there exists a $(\delta_1,
\delta_2)$-tester for $\mathcal{F}$. These definitions can be
extended to several classes of CPSOs.

The study of self-testing programs is a well-established research
area which was initiated by the work of Blum, Luby and
Rubinfeld\cite{BLR93}, Rubinfeld\cite{Rub90},
Lipton\cite{Lip91}
and Gemmel
\etal\cite{GLR91}. The purpose of a self-tester for a function
family is to detect by simple means if a program which is
accessible as an oracle computes a function from the given family.
This clearly inspired the definition of our self-testers
which have the particularity that they should test quantum objects
that
they can access only in some particular way. The analogy
with self-testing does not stop with the definition. One of the
main tools in self-testing of function families is the
characterization of these families by robust properties.
Informally, a property is robust if whenever a function satisfies
the property approximately, then it is close to a function which
satisfies it exactly. The concept of robustness was introduced and
its implication for self-testing was first studied by Rubinfeld and
Sudan\cite{RS96} and by Rubinfeld\cite{Rub94}. It will play a
crucial role in our case.

We note in the Preliminaries that for any real $\phi$ the states
$\ket{1}$ and $e^{i\phi}\ket{1}$ are experimentally
indistinguishable. This implies that if we start by only
distinguishing the classical states $0$ and $1$ then there are
families of CPSOs which are indistinguishable
as well.
For example, let
$\super{H}$ be the well-known Hadamard gate, and let
$\super{H}_{\phi}$ be the same gate expressed in the basis
$(\ket{0}, e^{i \phi}\ket{1})$, for $\phi \in [0,2\pi)$.
Any experiment that starts in state $0$ or $1$ and uses only
$\super{H}$ will produce outcomes $0$ and $1$ with the same
probabilities as the same experiment with $\super{H_{\phi}}$.
Thus no experiment that uses this quantum gate alone can
distinguish it from all the other Hadamard gates. Indeed,
as stated later in Fact~\ref{conjugate_basis}, a family
$\mathcal{F}$ containing $\super{H}$ can only be tested if the
entire Hadamard family $\mathcal{H}=\{\super{H}_\phi:\phi\in
[0,2\pi)\}$ is included in $\mathcal{F}$.

The main result of this paper is \textbf{\reftheo{maintheo}} which states
that for several sets of unitary CPSOs,
in particular,
the Hadamard gates
family, Hadamard gates together with $\cnot$
gates, and
Hadamard gates with $\cnot$
and phase rotation gates of angle $\pm\pi/4$, are testable.
This last family is of particular importance since
every triplet in the family
forms a universal and fault-tolerant set of gates
for quantum computation\cite{bmprv99}.

For the proof we will define the notion of experimental equations
which are functional equations for CPSOs corresponding to the
properties of the quantum gate that a self-tester can approximately
test. These tests are done via the interaction with the
experimental oracle. The proof itself contains three parts. In
\textbf{Theorems \ref{general}},
\textbf{\ref{couple}}, and \textbf{\ref{cnot}}  we will exhibit
experimental equations for the families of unitary CPSOs we want to
characterize. In \textbf{\reftheo{robustness}} we will show that
actually all experimental equations are robust;
in fact, the distance of a CPSO from
the target family is polynomially related to
the error tolerated in the experimental equations.
Finally
\textbf{\reftheo{robtotester}} gives self-testers for CPSO families
which are characterized by a finite set of robust experimental
equations.

In some cases, we are able to calculate explicitly the polynomial bound
in the robustness of experimental equations.
Such a result will be illustrated in \textbf{\reftheo{hadamardrob}}
for the equations characterizing the Hadamard family $\mathcal{H}$.

Technically, these results will be based on the representation of
one-qubit states and CPSOs in $\R^3$, where they are respectively
vectors in the unit ball of $\R^3$, and particular affine
transformations. This correspondence is known as the Bloch Ball
representation.

\section{Preliminaries}
\subsection{The quantum state}
A {\em pure state} in a quantum physical system is described by a
unit vector in a Hilbert space. In the {\em Dirac} notation it is
denoted by $\ket{\psi}$. In particular a {\em qubit} (a quantum
two-state system) is an element of the Hilbert space $\Hi$. The
orthonormal basis containing $\ket{0}$ and $\ket{1}$ is called the
{\em computational basis} of $\Hi$. Therefore a pure state
$\ket{\psi}\in\Hi$ is a {\em superposition} of the computational
basis states, that is, $\ket{\psi}=c_{0}\ket{0}+c_{1}\ket{1}$, with
$\abs{c_{0}}^{2}+\abs{c_{1}}^{2}=1$. A physical system which deals
with $n$ qubits is described mathematically by the
$2^n$-dimensional Hilbert space which is by definition
$\Hi\otimes\cdots\otimes\Hi$, that is, the $n^{\mbox{\footnotesize
th}}$ tensor power of $\Hi$. Let $N=2^n$. The computational basis
of this space consists of the $N$ orthonormal states $\ket{i}$ for
$0\leq i<N$. If $i$ is in binary notation $i_1 i_2\ldots i_n$, then
$\ket{i_1\ldots i_n}=\ket{i_1}\ldots\ket{i_n}$, where this is a
short notation for $\ket{i_1}\otimes\cdots\otimes\ket{i_n}$. All
vectors and matrices will be expressed in the computational basis.
The transposed complex conjugate $\ket{\psi}^\dag$ of $\ket{\psi}$
is denoted by $\bra{\psi}$. The inner product between $\ket{\psi}$
and $\ket{\psi'}$ is denoted by $\braket{\psi}{\psi'}$, and their
outer product by $\ketbra{\psi}{\psi'}$.

Quantum systems can also be in more general states than what
can be described by pure states. The most general states are
\emph{mixed states}, described by
a probability distribution over pure states. Such a mixture can be
denoted by $\{(p_{k},\ket{\psi_{k}}) : k\in\N \}$, where the system
is in the pure state $\ket{\psi_{k}}$ with probability $p_{k}$.

Different mixtures (even different pure states $\ket{\psi}$) can
represent the same physical system. This notational redundancy can
be avoided if we use the formalism of the density matrices.
 A \emph{density matrix} that represents an $n$-qubit
state is an $N\times N$ Hermitian semi-positive matrix with trace
$1$. The pure state $\ket{\psi}$ in this representation is
described by the density matrix $\psi=\ketbra{\psi}{\psi}$, and a
mixture $\{(p_{k},\ket{\psi_{k}})
: \ k\in\N\}$ by the density matrix
$\psi=\sum_{k\in\N}p_{k}\ketbra{\psi_{k}}{\psi_{k}}$. For example,
the pure states $e^{i\gamma}\ket{\psi}$, for $\gamma\in[0,2\pi)$,
or the mixtures $\{(\half,\ket{0}),(\half,\ket{1})\}$ and
$\{(\half,\frac{\ket{0}+\ket{1}}{\sqrt{2}}),
(\half,\frac{\ket{0}-\ket{1}}{\sqrt{2}})\}$ have respectively the
same density matrix.

Since a density matrix is Hermitian semi-positive, its eigenvectors
are orthogonal and its eigenvalues are non-negative. Because
the density matrix has trace $1$, its eigenvalues sum to $1$.
Therefore a density matrix
represents the mixture of its orthonormal eigenvectors, where the
probabilities are the respective eigenvalues. Note that diagonal
density matrices correspond to a mixture over pure states
$\ket{i}$, for $0\leq i<N$. Density matrices that represent pure
states have a simple algebraic characterization: $\rho$ is a pure
state if and only if it has two eigenvalues, $0$ with multiplicity
$N-1$ and $1$ with multiplicity $1$, equivalently $\rho$ is a pure
state exactly when $\rho^2=\rho$.

A $2\times 2$ Hermitian matrix of unit trace is semi-positive if
and only if its determinant is between $0$ and $1/4$. Therefore in
the case of one qubit, any density matrix $\rho$ can be written  as
$\rho = p\ketbra{0}{0}+(1-p)\ketbra{1}{1}
+\alpha\ketbra{1}{0}+{\alpha^*}\ketbra{0}{1}$, where $p\in[0,1]$,
and $\alpha$ is a complex number such that $\abs{\alpha}^2\leq
p(1-p)$. This density matrix will be denoted by $\rho(p,\alpha)$.
Remark that $\rho(p,\alpha)$ is a pure state exactly when
$\abs{\alpha}^2= p(1-p)$, that is, its determinant is $0$.

\subsection{Superoperators}
The evolution of physical systems is described by specific
transformations on density matrices, that is, on operators. A
{\em superoperator} for $n$ qubits is a linear transformation on
$\C^{N\times N}$. A {\em positive} superoperator (PSO) is a
superoperator that sends density matrices to density matrices. A
{\em completely positive} superoperator (CPSO) $\super{G}$ is a PSO
such that for all positive integers $M$,
 $\super{G}\otimes\super{I}_{M}$ is also a PSO,
where $\super{I}_{M}$ is the identity on $\C^{M\times M}$.
CPSOs are exactly the physically allowed transformations on
density matrices. An example of a PSO for one qubit that is not a
CPSO is the transpose superoperator $\super{T}$ defined by
$\super{T}(\ketbra{i}{j})=\ketbra{j}{i}$, for $0\leq i,j\leq 1$.

Quantum~computation is based on the possibility of constructing
some particular CPSOs, {\em unitary} superoperators, which preserve
the set of pure states. These operators are characterized by
transformations from $\U(N)$, the set of $N\times N$ unitary
matrices. For any $A\in \U(N)$, we define a CPSO which maps a
density matrix $\rho$ into $A\rho A^\dag$. When the underlying
unitary transformation $A$ is clear from the context, by somewhat
abusing the notation, we will denote this CPSO simply by
$\super{A}$. If $\ket{\psi'}$ denotes $A\ket{\psi}$, then the
unitary superoperator $\super{A}$ maps the pure state $\psi$ to the
pure state $\psi'$. As was the case in the Dirac representation of
states, there is the same phase redundancy in the set of unitary
transformations $\U(N)$. If $A\in\U(N)$, then for all $\gamma\in
[0,2\pi)$, the transformations $e^{i\gamma} A$ are different,
however the corresponding superoperators are identical. We will
therefore focus on $\U(N)/\U(1)$.

\subsection{Measurements}
Measurements form another important class of (non-unitary) CPSOs.
They describe physical transformations corresponding to the
observation of the system. We will define now formally one of the
simplest classes of measurements which correspond to the
projections to elements of the computational basis.

A {\em Von Neumann measurement in the computational basis} of $n$
qubits is the $n$-qubit CPSO $\super{M}$ that, for every
density matrix
$\rho$, satisfies $\super{M}(\rho)_{i,i}=\rho_{i,i}$ and
$\super{M}(\rho)_{i,j}=0$, for $i\neq j$.

In the case of one qubit, the Von Neumann measurement in the
computational basis maps the density matrix $\rho(p,\alpha)$ into
$\rho(p,0)$. We will say that $p=\bra{0}\rho\ket{0}$ is the {\em
probability of measuring} $\ketbra{0}{0}$, and we will denote it by
$\PM{0}{\rho}$.

In general, a {\em Von Neumann measurement} of $n$ qubits in any
basis can be viewed as the Von Neumann measurement in the
computational basis preceded by some unitary superoperator.

\subsection{The Bloch Ball representation}

Specific for the one-qubit case, there is an isomorphism
between the group $\U(2)/\U(1)$ and the special rotation group $\SO(3)$,
the set of $3\times 3$ orthogonal matrices with determinant $1$.
This allows us to
represent one-qubit states as vectors in the unit ball of $\R^3$,
and unitary superoperators as rotations on $\R^3$.
We will now describe exactly this correspondence.

The {\em Bloch Ball} $\BB$ (respectively {\em Bloch Sphere} $\BS$)
is the unit ball (respectively unit sphere) of the Euclidean affine space
$\R^{3}$.
Any point $\ball{u}\in\R^{3}$ determines a vector
with the same coordinates which we will also denote by $\ball{u}$.
The inner product of $\ball{u}$ and $\ball{v}$ will be denoted by
$(\ball{u},\ball{v})$, and
their Euclidean norm by $\norm{\ball{u}}$.

Each point $\ball{u}\in\R^{3}$ can be also characterized by its
norm $r\geq 0$,
its latitude $\theta\in[0,\pi]$, and its longitude
$\phi\in[0,2\pi)$.
The {\em latitude} is the angle between the $z$-axis and the vector
$\ball{u}$, and
the {\em longitude} is the angle between the $x$-axis and the orthogonal
projection of $\ball{u}$ in the plane defined by $z=0$.
If $\ball{u}=(x,y,z)$, then
these parameters satisfy
$x=r\sin\theta\cos\phi$,
$y=r\sin\theta\sin\phi$ and $z=r\cos\theta$.

For every density matrix $\rho$ for one qubit there exists a
unique point
$\ball{\rho}=(x,y,z) \in\BB$ such that
\begin{eqnarray*}
\rho
&=&\frac{1}{2}\dens{1+z}{x-i y}{x+i y}{1-z}.
\end{eqnarray*}
This mapping is a bijection that also obeys
\begin{eqnarray*}
\ball{\rho(p,\alpha)} & = & (2\Re(\alpha),2\Im(\alpha),2p-1).
\end{eqnarray*}

In this formalism, the pure states are nicely characterized in $\BB$
by their norm.
\begin{fact}
A density matrix $\rho$ represents a pure state if and only if
$\ball{\rho}\in\BS$, that is, $\norm{\ball{\rho}}=1$.
\end{fact}
Also, if $\theta\in[0,\pi]$ and $\phi\in[0,2\pi)$ are respectively
the latitude and the longitude of $\ball{\psi}\in\BS$, then the
corresponding density matrix represents a pure state and satisfies
$\ket{\psi} =
\cos(\theta/2)\ket{0}+\sin(\theta/2)e^{i\phi}\ket{1}$.
Observe that the pure states $\ket{\psi}$ and $\ket{\psi^{\perp}}$
are orthogonal if and only if $\ball{\psi}=-\ball{\psi^\perp}$. We
will use the following notation for the six pure states along the
$x$, $y$ and $z$ axes: $\ket{\zeta_x^\pm} =
\halfsqrt(\ket{0}\pm\ket{1})$, $\ket{\zeta_y^\pm} =
\halfsqrt(\ket{0}\pm i \ket{1})$, $\ket{\zeta_z^+} = \ket{0}$, and
$\ket{\zeta_z^-} = \ket{1}$, with the respective coordinates $(\pm
1,0,0)$, $(0,\pm 1,0)$ and $(0,0,\pm 1)$ in $\R^3$.

For each CPSO $\super{G}$, there exists a unique affine
transformation $\affine{G}$ over $\R^{3}$, which
maps the ball $\BB$ into $\BB$ and
is such that, for all density matrices $\rho$,
$\affine{G}(\ball{\rho})=\ball{\super{G}(\rho)}$.
Unitary superoperators have a nice characterization
in $\BB$.
\begin{fact}
The map between $\U(2)/\U(1)$ and $\SO(3)$, which sends $A$ to
$\affine{A}$, is an isomorphism.
\end{fact}
For $\alpha\in(-\pi,\pi]$, $\theta\in[0,\frac{\pi}{2}]$, and
$\phi\in[0,2\pi)$, we will define the unitary transformation
$R_{\alpha,\theta,\phi}$ over $\C^{2}$. If
$\ket{\psi}=\cos(\theta/2)\ket{0} +e^{i\phi}\sin(\theta/2)\ket{1}$
and $\ket{\psi^\perp}=\sin(\theta/2)\ket{0}
-e^{i\phi}\cos(\theta/2)\ket{1}$
then by definition $R_{\alpha,\theta,\phi}\ket{\psi}=\ket{\psi}$
and
$R_{\alpha,\theta,\phi}\ket{\psi^\perp}=e^{i\alpha}\ket{\psi^\perp}$.
If $\super{A}$ is a unitary superoperator then we have
$\super{A}=\super{R}_{\alpha,\theta,\phi}$ for some $\alpha$,
$\theta$, and $\phi$. In $\R^3$ the transformation
$\affine{R}_{\alpha,\theta,\phi}$ is the rotation of angle $\alpha$
whose axis cuts the sphere $\BS$ in the points $\ball{\psi}$ and
$\ball{\psi^\perp}$. Note that for $\theta=0$ the CPSO
$\super{R}_{\alpha,0,\phi}$ does not depend on $\phi$. We will
denote this phase rotation by $\super{R}_{\alpha}$.

The affine transformation in $\BB$
which corresponds to the Von Neumann measurement in the computational basis
is the
orthogonal projection to the $z$-axis.
Therefore it maps
$\ball{\rho}=(x,y,z)$ into $(0,0,z)$, the point which corresponds to the
density matrix $\frac{1+z}{2}\ketbra{0}{0}+\frac{1-z}{2}\ketbra{1}{1}$.
Thus $\PM{0}{\rho}=\frac{1+z}{2}$.

\subsection{Norm and distance}
Let $N=2^n$. We will consider the {\em trace norm} on $\C^{N\times
N}$ which is defined as follows: for all $V\in\C^{N\times
N}$,
$\normone{V}=\Tr\sqrt{V^{\dag}V}$. This norm has several advantages
when we consider the difference of density matrices. Given a Von
Neumann measurement, a density matrix induces a probability
distribution over the basis of the measurement. The trace norm of
the difference of two density matrices is the maximal variation
distance between the two induced probability distributions, over
all Von Neumann measurements. It also satisfies the following
properties.
\begin{fact}
For all density matrices $\rho(p,\alpha)$ and $\rho(q,\beta)$ for
one qubit we have:
\begin{equation*}
\begin{array}{rcccl}
\normone{\rho(p,\alpha)-\rho(q,\beta)}&=&
\norm{\ball{\rho(p,\alpha)}-\ball{\rho(q,\beta)}}
&=&2\sqrt{(p-q)^2+\abs{\alpha-\beta}^2}.
\end{array}
\end{equation*}
\end{fact}
\begin{fact}
For all $V \in\C^{N \times N}$ and $W\in\C^{M\times M}$ we have
$\normone{V\ox W}=\normone{V}\normone{W}$ and $\abs{\Tr(V)}\leq\normone{V}$.
For density matrices $\rho$ it holds that $\normone{\rho}=1$.
\end{fact}
For $n$-qubit superoperators, the superoperator norm
associated to the trace norm is defined as
\begin{eqnarray*}
\normso{\super{G}}& = &
\sup\{
\normone{\super{G}(V)} : \normone{V}=1\}.
\end{eqnarray*}
This norm is always $1$ when $\super{G}$ is a CPSO.
The norm $\normso{\,}$ can be easily generalized for $k$-tuples
of superoperators by
$\normso{(\super{G}_1,\ldots,\super{G}_k)}=
\max(\normso{\super{G}_1},\ldots,\normso{\super{G}_k}).$
We will denote by $\distso$ the natural induced distance
by the norm $\normso{\,}$.

\section{Properties of CPSOs}
Here we will establish the properties of CPSOs that we will need
for the characterization of our CPSO families.
In this extended abstract we will
omit the proof of Lemma \ref{lemma1}, and the proof of Lemma
\ref{lemma2} will be in \refap{APL2}.

\begin{lemma}\label{lemma1}
Let $\super{G}$ be a CPSO for one qubit,
and let $\rho$ and $\tau$ be density
matrices for one qubit.
\begin{enumerate}
\item[(a)]
$\normone{\super{G}(\rho)-\super{G}(\tau)}\leq\normone{\rho-\tau}$.
\item[(b)] If $\super{G}$ is not constant and $\super{G}(\rho)$ is a pure
state
then $\rho$ is a pure state.
\end{enumerate}
\end{lemma}

An affine transformation of $\R^3$ is uniquely defined by the images
of four non-coplanar points. Surprisingly, if the transformation is
a CPSO for one qubit, the images of three points are
sometimes sufficient. The following will make this precise more generally
for $n$ qubits.
\begin{lemma}\label{lemma2}
Let $n\geq 1$ be an integer, and
let $\rho_1$, $\rho_2$, and $\rho_3$ be three distinct one-qubit
density matrices representing pure states,
such that the plane in $\R^3$
containing the points $\ball{\rho_1},\ball{\rho_2},\ball{\rho_3}$
goes through the center of $\BB$. If $\super{G}$ is a CPSO for
$n$ qubits which acts as the identity on the set
$\{\rho_1,\rho_2,\rho_3\}^{\otimes n}$,
then $\super{G}$ is the identity mapping.
\end{lemma}

We also use the property that for CPSOs unitarity and invertibility
are equivalent (see e.g. \cite[Ch. 3, Sec. 8]{Pre98}).
\begin{lemma} \label{lemma1c}
Let $\super{G}$ be a CPSO for $n$ qubits.
If there exists a CPSO $\super{H}$ for $n$ qubits
such that $\super{H}\circ\super{G}$ is the
identity mapping, then $\super{G}$ is a unitary superoperator.
\end{lemma}

\section{Characterization}
\subsection{One-Qubit CPSO Families}
In this section, every CPSO will be for one qubit.
First we define the notion of experimental
equations, and then we show that several important
CPSO families
are characterizable by them.

An {\em experimental equation} in one variable
is a CPSO equation of the form
\begin{eqnarray}\label{expCPSOeq}
\PM{0}{\super{G}^{k}(\ketbra{b}{b})} & = & r,
\end{eqnarray}
where $k$ is a non-negative integer, $b\in\{0,1\}$, and $0\leq
r\leq 1$. We will call the left-hand side of the equation the {\em
probability term}, and the right-hand side the {\em constant term}.
The {\em size} of this equation is $k$. A CPSO $\super{G}$ will
``almost'' satisfy the equations if, for example, it is the result
of adding small systematic and random errors (independent of time)
to a CPSO that does.
 For $\eps\geq 0$, the CPSO
$\super{G}$ {\em $\eps$-satisfies} \refeq{expCPSOeq} if
$\abs{\PM{0}{\super{G}^{k} (\ketbra{b}{b})}-r}\leq\eps,$ and when
$\eps=0$ we will just say that $\super{G}$ {\em satisfies}
\refeq{expCPSOeq}. Let $(E)$ be a finite set of experimental
equations. If $\super{G}$ $\eps$-satisfies all equations in $(E)$
we say that $\super{G}$ $\eps$-satisfies $(E)$. If some $\super{G}$
satisfies $(E)$ then $(E)$ is {\em satisfiable}. The set
$\{\super{G} :
\super{G}\mbox{ satisfies $(E)$}\}$
will be denoted by $\mathcal{F}_{(E)}$.
A family $\mathcal{F}$ of CPSOs is {\em characterizable}
if it is $\mathcal{F}_{(E)}$
for some finite set $(E)$ of experimental equations.
In this case we say that $(E)$ {\em characterizes} $\mathcal{F}$.

All these definitions
generalize naturally for $m$-tuples of CPSOs for
$m\geq 2$.
In what follows we will need
only the case $m=2$.
An {\em experimental equation} in two CPSO variables
is an equation of the form
\begin{eqnarray*}\label{expCPSOeq2}
\PM{0}{\super{F}^{k_{1}}\circ\super{G}^{l_1}\circ\cdots\circ
\super{F}^{k_{t}}\circ\super{G}^{l_{t}}(\ketbra{b}{b})} & = & r,
\end{eqnarray*}
where $k_1,\ldots,k_t,l_1,\ldots,l_t$ are non-negative integers,
$b\in\{0,1\}$, and $0\leq r\leq 1$.

We discuss now the existence of finite sets of experimental
equations in one variable that characterize unitary superoperators,
that is, the operators $\super{R}_{\alpha,\theta,\phi}$, for
$\alpha\in(-\pi,\pi]$, $\theta\in[0,\pi/2]$, and $\phi\in[0,2\pi)$.
First observe that due to the restrictions of experimental
equations, there are unitary superoperators that they cannot
distinguish.
\begin{fact}  \label{conjugate_basis}
Let $\alpha\in[0,\pi]$, $\theta\in[0,\pi/2]$, and
$\phi_1,\phi_2\in[0,2\pi)$ such that $\phi_1\neq\phi_2$.
Let $(E)$ be a finite set of experimental
equations in $m$ variables. If $(\super{R}_{\alpha,\theta,\phi_1},
\super{G}_2,\ldots,\super{G}_m)$ satisfies $(E)$ then
there exist $\super{G}'_2,\ldots,\super{G}'_m$ and
$\super{G}''_2,\ldots,\super{G}''_m$ such that
$(\super{R}_{-\alpha,\theta,\phi_1},
\super{G}'_2,\ldots,\super{G}'_m)$
and $(\super{R}_{\alpha,\theta,\phi_2},
\super{G}''_2,\ldots, \super{G}''_m)$ both satisfy $(E)$.
\end{fact}
In the Bloch Ball formalism this corresponds to the following
degrees of freedom in the choice of
the orthonormal basis of $\R^3$.
Since
experimental
equations contain exactly the states $\ketbra{0}{0}$ and
$\ketbra{1}{1}$
there is no freedom in the choice of the $z$-axis, but there
is complete freedom in the choice of the $x$ and $y$ axes.
The indistinguishability of the latitude $\phi$ corresponds
to the freedom of choosing the oriented $x$-axis, and the
indistinguishability of the sign of $\alpha$ corresponds
to the freedom of choosing the orientation of the $y$-axis.

We introduce the following notations.
Let $\mathcal{R}_{\alpha,\theta}$ denote the superoperator family
$\{\super{R}_{\pm\alpha,\theta,\phi} : \phi\in[0,2\pi)\}$.
For  $\phi\in[0,2\pi)$,
let the $\neg_\phi$ transformation be defined by
$\neg_\phi\ket{0} = e^{i\phi}\ket{1}$
and $\neg_\phi(e^{i\phi}\ket{1})=\ket{0}$,
and recall that the Hadamard transformation
$H_\phi$ obeys $H_{\phi}\ket{0}=(\ket{0}+e^{i\phi}\ket{1})/\sqrt{2}$
and $H_{\phi}(e^{i\phi}\ket{1})=(\ket{0}-e^{i\phi}\ket{1})/\sqrt{2}$.
Observe that $\super{H}_\phi=\super{R}_{\pi,\pi/4,\phi}$
and $\super{\neg}_\phi=\super{R}_{\pi,\pi/2,\phi}$, for
$\phi\in[0,2\pi)$.
Finally let $\mathcal{H}=
\{\super{H}_\phi : \phi\in[0,2\pi)\}$, and
$\mathcal{N}=\{\super{\neg}_\phi : \phi\in[0,2\pi)\}$.

Since the sign of $\alpha$ cannot be determined, we will assume
that $\alpha$ is in the interval $[0,\pi]$. We will also consider
only unitary superoperators such that $\alpha/\pi$ is rational.
This is a reasonable choice since these superoperators form a dense
subset of all unitary superoperators. For such a unitary
superoperator, let $n_\alpha$ be the smallest positive integer $n$
for which $n\alpha=0 \mod{2\pi}$. Then either $n_\alpha=1$, or
$n_\alpha\geq 2$ and there exists $t\geq 1$ which is coprime with
$n_\alpha$ such that $\alpha=(t/n_\alpha)2\pi$. Observe that the
case $n_\alpha=1$ corresponds to the identity superoperator.

Our first theorem shows that almost all families
$\mathcal{R}_{\alpha,\theta}$ are characterizable by
some finite set of experimental equations.
In particular $\mathcal{H}$ is characterizable.
\begin{theorem}\label{general}
Let $(\alpha,\theta)\in(0,\pi]\times(0,\pi/2]\backslash\{(\pi,\pi/2)\}$
be such that
$\alpha/\pi$ is rational.
Let $z_k(\alpha,\theta)=\cos^{2}\theta+\sin^{2}\theta\cos(k\alpha)$.
Then the following experimental
equations characterize $\mathcal{R}_{\alpha,\theta}$:
\begin{equation*}\label{generaltest1}
\begin{array}{rcll}
\PM{0}{\super{G}^{n_\alpha}(\ketbra{1}{1})}=  0
& \textrm{ and } &
\PM{0}{\super{G}^{k}(\ketbra{0}{0})} =  \half+\half z_{k}(\alpha,\theta),
& k\in\positive{n_\alpha}.
\end{array}
\end{equation*}
\end{theorem}
\proof
First observe that every CPSO in $\mathcal{R}_{\alpha,\theta}$
satisfies the equations of the theorem since the $z$-coordinate of
$\ball{\super{R}_{\alpha,\theta,\phi}^k(\ketbra{0}{0})}$ is
$z_k(\alpha,\theta)$ for every $\phi\in[0,2\pi)$. Let $\super{G}$
be a CPSO which
 satisfies these equations. We will prove that
$\super{G}$ is a unitary superoperator. Then,
\reffact{unicityofangles} implies that
$\super{G}\in\mathcal{R}_{\alpha,\theta}$.

Since $z_1(\alpha,\theta)\neq \pm 1$, we know
$\super{G}(\ketbra{0}{0})\not\in\{\ketbra{0}{0},\ketbra{1}{1}\}$.
Observing that $\super{G}^{n_\alpha}(\ketbra{0}{0})=\ketbra{0}{0}$,
\reflemma{lemma1}(b) implies that $\super{G}(\ketbra{0}{0})$ is a
pure state. Thus $\ketbra{0}{0}$, $\ketbra{1}{1}$, and
$\super{G}(\ketbra{0}{0})$ are distinct pure states, and since
$\super{G}^{n_\alpha}$ acts as the identity on them, by
\reflemma{lemma2} it is the identity mapping. Hence by
\reflemma{lemma1c} $\super{G}$ is a unitary superoperator.
\qed
\begin{fact}\label{unicityofangles}
Let $\alpha\in(0,\pi]$,
$\theta\in(0,\pi/2]$,
$\alpha'\in(-\pi,\pi]$, $\theta'\in(0,\pi/2]$
be such that $\alpha/\pi$ is rational.
If $z_k(\alpha,\theta)=
z_k(\alpha',\theta')$, for
$k\in\positive{n_\alpha}$, then $\abs{\alpha'}=\alpha$
and $\theta'=\theta$.
\end{fact}

The remaining families $\mathcal{R}_{\alpha,\theta}$ for which
$\alpha/\pi$ is rational are
$\{\super{R}_{-\alpha},\super{R}_\alpha\}$, for $\alpha\in[0,\pi]$,
and $\mathcal{N}$. Let us recall that $\super{M}$ is the CPSO which
represents the Von Neumann measurement in the computational basis.
Since $\super{M}$ satisfies exactly the same equations as
$\super{R}_{\pm\alpha}$, and $\super{\neg}_0\circ\super{M}$
satisfies exactly the same equations as $\super{\neg}_\phi$, for
any $\phi\in[0,2\pi)$, these families are not characterizable by
experimental equations in one variable. Nevertheless it turns out
that together with the family $\mathcal{H}$ they become
characterizable. This is stated in the following theorem whose
proof is omitted.

\begin{theorem}~\label{couple}
The family
$\{(\super{H}_\phi,\super{\neg}_{\phi}) : \phi\in[0,2\pi)\}
\subset \mathcal{H}\times\mathcal{N}$  is characterized by
the experimental equations in two variables $(\super{F},\super{G})$:
\begin{equation*}
\left\{{\begin{array}{lll}
\PM{0}{\super{F}(\ketbra{0}{0})}=\half,\!\!&
\PM{0}{\super{F}^2(\ketbra{0}{0})}=1,\!\!&
\PM{0}{\super{F}^2(\ketbra{1}{1})}=0,\!\!\\
\vspace{-.1cm}
& & \\
\PM{0}{\super{G}(\ketbra{0}{0})}=0,\!\!&
\PM{0}{\super{G}(\ketbra{1}{1})}=1,\!\!& \\
\vspace{-.1cm}
& & \\
\PM{0}{\super{F}\circ \super{G}^2\circ \super{F}(\ketbra{0}{0})}=1,\!\!&
\PM{0}{\super{F}\circ \super{G}\circ \super{F}(\ketbra{0}{0})}=1.&\!\!\\
\end{array}}\right.
\end{equation*}
If $\alpha/\pi$ is rational, then the family
$\mathcal{H}\times\{ \super{R}_{\pm\alpha}\}$
is characterized by the experimental equations
in two variables $(\super{F},\super{G})$:
\begin{equation*}
\left\{{\begin{array}{lll}
\PM{0}{\super{F}(\ketbra{0}{0})}=\half,\!\!&
\PM{0}{\super{F}^2(\ketbra{0}{0})}=1,\!\!&
\PM{0}{\super{F}^2(\ketbra{1}{1})}=0,\!\!\\
\vspace{-.1cm}
& & \\
\PM{0}{\super{G}(\ketbra{0}{0})}=1,\!\!&
\PM{0}{\super{G}(\ketbra{1}{1})}=0,\!\!& \\
\vspace{-.1cm}
& & \\
\PM{0}{\super{F}\circ \super{G}^{n_\alpha}\circ
\super{F}(\ketbra{0}{0})}=1,\!\!&
\PM{0}{\super{F}\circ \super{G}\circ
\super{F}(\ketbra{0}{0})}=\half+\half\cos\alpha.\!\!& \\
\end{array}}\right.
\end{equation*}
\end{theorem}

\subsection{Characterization of $\super{\cnot}$ gates}
In this section we will extend our theory of characterization
of CPSO families for several qubits.
In particular, we will show that the family of $\super{\cnot}$ gates
together with
the family $\mathcal{H}$ is characterizable.
First we need some definitions.

For every $\phi\in[0,2\pi)$, we define $\cnot_\phi$ as
the only unitary transformation over $\C^4$ satisfying
$\cnot_{\phi}(\ket{0}\ket{\psi})=\ket{0}\ket{\psi}$
and $\cnot_{\phi}\ket{1}\ket{\psi}=\ket{1}\neg_\phi\ket{\psi}$, for
all $\ket{\psi}\in\C^2$.

We extend the definition of the experimental equation for CPSOs
given in~\refeq{expCPSOeq2} for $n$ qubits.
It is an equation of the form
\begin{eqnarray}\label{expCPSOeq3}
\PM{v}{\super{F}^{k_{1}}\circ\super{G}^{l_1}\circ\cdots\circ
\super{F}^{k_{t}}\circ\super{G}^{l_{t}}(\ketbra{w}{w})}& = & r,
\end{eqnarray}
where in addition to the notation of~\refeq{expCPSOeq2}
$v,w\in\{0,1\}^n$, and $\mathrm{Pr}^{v}$ is the probability of
measuring $\ketbra{v}{v}$. For the variables $\super{F}$ and
$\super{G}$ of~\refeq{expCPSOeq3}, we also allow both the tensor
product of two CPSO variables and the tensor product of a CPSO
variable with the identity. We now state the characterization.
\begin{theorem}~\label{cnot}
The family
$\{(\super{H}_\phi,\super{\cnot}_{\phi}) : \phi\in[0,2\pi)\}$
is characterized by
the experimental equations in two variables $(\super{F},\super{G})$:
\begin{equation*}
\left\{{\begin{array}{llll}
\PM{0}{\super{F}(\ketbra{0}{0})}=\half,\!\!&
\PM{0}{\super{F}^2(\ketbra{0}{0})}=1,\!\!&
\PM{0}{\super{F}^2(\ketbra{1}{1})}=0,\!\!& \\
\vspace{-.1cm}
&&& \\
\PM{00}{\super{G}(\ketbra{00}{00})}=1,\!\!&
\PM{01}{\super{G}(\ketbra{01}{01})}=1,\!\!&
\PM{11}{\super{G}(\ketbra{10}{10})}=1,\!\!&
\PM{10}{\super{G}(\ketbra{11}{11})}=1,\!\!\\
\vspace{-.1cm}
&&& \\
\multicolumn{2}{l}{
\PM{00}{(\super{I}_2\otimes \super{F})\circ \super{G}\circ
(\super{I}_2\otimes \super{F})(\ketbra{00}{00})}=1,\!\!}&
\multicolumn{2}{l}{
\PM{10}{(\super{I}_2\otimes \super{F})\circ \super{G}\circ
(\super{I}_2\otimes \super{F})(\ketbra{10}{10})}=1,\!\!}\\
\vspace{-.1cm}
&&& \\
\multicolumn{2}{l}{
\PM{00}{(\super{F}\otimes\super{I}_2)\circ \super{G}^2\circ
(\super{F}\otimes\super{I}_2)(\ketbra{00}{00})}=1,\!\!}&
\multicolumn{2}{l}{
\PM{01}{(\super{F}\otimes\super{I}_2)\circ \super{G}^2\circ
(\super{F}\otimes\super{I}_2)(\ketbra{01}{01})}=1,\!\!}\\
\vspace{-.1cm}
&&& \\
\multicolumn{2}{l}{
\PM{00}{(\super{F}\otimes\super{F})\circ \super{G}\circ
(\super{F}\otimes\super{F})(\ketbra{00}{00})}=1.\!\!}&&
\end{array}}\right.
\end{equation*}
\end{theorem}
\proof
Let $\super{F}$ and $\super{G}$ satisfy these equations.
By \reftheo{general}, with $\alpha=\pi$
and $\theta=\pi/4$, the first three equations imply that
$\super{F}=\super{H}_\phi$, for some $\phi\in[0,2\pi)$.
Using \reflemma{lemma2}, the remaining equations imply
that $\super{G}^2=\super{I}_4$,
and it follows from \reflemma{lemma1c}
that $\super{G}$ is a unitary CPSO. A straightforward
verification then shows that indeed $\super{G}=\super{\cnot_\phi}$.
\qed

\section{Robustness}
In this section we introduce the notion of robustness for
experimental equations which will be the crucial ingredient for
proving self-testability. In this extended abstract we will deal
only with the case of experimental equations for one qubit and in
one variable. {From} now on $(E)$ will always denote a set of such
equations. Similar results can be obtained for several qubits and
several variables.

\begin{definition}
Let $\eps,\delta\geq 0$, and let $(E)$ be
a finite satisfiable set of experimental equations.
We say that $(E)$ is {\em ($\eps,\delta$)-robust}
if whenever a CPSO $\super{G}$
$\eps$-satisfies $(E)$, we have
$\distso(\super{G},\mathcal{F}_{(E)})\leq\delta$.
\end{definition}
When a CPSO family is characterized by a
finite set of experimental equations $(E)$, one would like to prove
that $(E)$ is robust. The next theorem shows that this is always
the case.
\begin{theorem}\label{robustness}
Let $(E)$ be a finite satisfiable set of experimental
equations. Then there exists an integer $k\geq 1$ and a real $C>0$
such that for all $\eps\geq 0$,
$(E)$ is ($\eps,C\eps^{1/k}$)-robust.
\end{theorem}
\proof
We will use basic notions from algebraic geometry for which we refer
the reader for example to~\cite{br90}.
In the proof, $\C$ is identified with $\R^2$. Then the set $K$ of
CPSOs for one qubit is a real compact semi-algebraic set. Suppose
that in $(E)$ there are $d$ equations. Let $f:K\rightarrow\R$ be
the function that maps the CPSO $\super{G}$ to the maximum of the
magnitudes of the difference between the probability term and the
constant term of the $i^{\mbox{\scriptsize th}}$ equation in
$(E)$, for $i=1,\ldots,d$. By definition of $f$, we get
$f^{-1}(0)=\mathcal{F}_{(E)}$. Moreover, $f$ is a continuous
semi-algebraic function, since it is the maximum of the magnitudes
of polynomial functions in the (real) coefficients  of $\super{G}$.

Let $g:K\rightarrow\R$ defined in $\super{G}$ by
$g(\super{G})=\distso(\super{G},\mathcal{F}_{(E)})$. Since $K$ is a
compact semi-algebraic set, $g$ is a continuous semi-algebraic
function. Moreover, for all $\super{G}\in K$, we have
$f(\super{G})=0$ if and only if $g(\super{G})=0$. Then
\reffact{semi-alg} concludes the proof.
\qed

For a proof of the following fact,
see for example~\cite[Prop.~2.3.11]{br90}.
\begin{fact}[Lojasiewicz's inequality]\label{semi-alg}
Let $X\subseteq\R^m$ be a compact semi-algebraic set.
Let $f,g:X\rightarrow\R$ be continuous semi-algebraic functions.
Assume that for all $x\in X$, if
$f(x)=0$ then $g(x)=0$.
Then there exists an integer $k\geq 1$ and a real $C>0$ such that,
for all $x\in X$, $\abs{g(x)}^k\leq C\abs{f(x)}$.
\end{fact}

In some cases we can explicitly compute the constants $C$ and $k$
of \reftheo{robustness}.
We will illustrate these techniques with the equations in
\reftheo{general} for the case $\alpha=\pi$ and $\theta=\pi/4$. Let
us recall that these equations characterize the set $\mathcal{H}$.
\begin{theorem}\label{hadamardrob}
For every $0 \leq \eps\leq 1$, the following equations are
($\eps,4579\sqrt{\eps}$)-robust:
\begin{eqnarray*}\label{approxhadamard}
\PM{0}{\super{G}(\ketbra{0}{0})}=\half,&
\PM{0}{\super{G}^2(\ketbra{0}{0})}=1,
\textrm{ and} &
\PM{0}{\super{G}^2(\ketbra{1}{1})}=0.
\end{eqnarray*}
\end{theorem}
The proof of this theorem will be given in \refap{APTH}.

\section{Quantum Self-Testers}
In this final section we define formally our testers
and establish the relationship between robust equations
and testability.
Again, we will do it here only for the case of one qubit and one
variable.
Let $\super{G}$ be a CPSO.
The {\em experimental oracle} $\oracle{\super{G}}$ for $\super{G}$
is a probabilistic procedure.
It takes inputs from $\{0,1\}\times\N$
and generates outcomes from the set $\{0,1\}$ such that
for every $k\in\N$,
\begin{eqnarray*}
\Pr[\oracle{\super{G}}(b,k)=0]
& = &
\PM{0}{\super{G}^{k}(\ketbra{b}{b})}.
\end{eqnarray*}
An oracle program $T$ with an experimental oracle
$\oracle{\super{G}}$ is a program denoted by
$T^{\oracle{\super{G}}}$ which can ask queries
from the experimental oracle
in the following sense:
when it presents a query $(b,k)$ to the oracle,
in one computational step it receives the probabilistic outcome of
$\oracle{\super{G}}$ on it.
\begin{definition}
Let $\mathcal{F}$ be a family of CPSOs, and
let $0\leq\delta_1\leq\delta_2<1$.
 A {\em $(\delta_1,\delta_2)$-tester for} $\mathcal{F}$
is a probabilistic oracle program $T$
such that for every CPSO $\super{G}$,
\begin{itemize}
\item if $\distso(\super{G},\mathcal{F})\leq\delta_1$
then
$\Pr[T^{\oracle{\super{G}}}\mbox{ says { \tt PASS}}]\geq 2/3$,
\item if $\distso(\super{G},\mathcal{F})>\delta_2$ then
$\Pr[T^{\oracle{\super{G}}}\mbox{ says { \tt FAIL}}]\geq 2/3$,
\end{itemize}
where the probability is taken over the probability distribution of the
outcomes
of the experimental oracle and the internal coin tosses of the program.
\end{definition}
\begin{theorem}\label{robtotester}
Let $\eps,\delta>0$, and let $(E)$ be a satisfiable set of $d$ experimental
equations such that the size of every equation is at most $k$.
If $(E)$ is ($\eps,\delta$)-robust then there exists an
$(\eps/(3k),\delta)$-tester for $\mathcal{F}_{(E)}$
which makes $O(d\ln (d)/\eps^2)$ queries.
\end{theorem}
\textsc{Sketch of proof.}
We will describe a probabilistic oracle program $T$. Let
$\super{G}$ be a CPSO. We can suppose that for every equation in
$(E)$, $T$ has a rational number $\tilde{r}$ such that
$\abs{\tilde{r}-r}\leq\eps/6$, where $r$ is the constant term of
the equation. By sampling the oracle $\oracle{\super{G}}$, for
every equation in $(E)$, $T$ obtains a value $\tilde{p}$ such that
$\abs{\tilde{p}-p}\leq\eps/6$ with probability at least $1-1/(3d)$,
where $p$ is the probability term of the equation. A standard
Chernoff bound argument shows that this is feasible with
$O(\ln(d)/\eps^2)$ queries for each equation. If for every equation
$\abs{\tilde{p}-\tilde{r}}\leq 2\eps/3$, then $T$ says {\tt PASS},
otherwise $T$ says {\tt FAIL}. Using the robustness of $(E)$ and
\reflemma{lemmar}, one can verify that $T$ is a
$(\eps/(3k),\delta)$-tester for $\mathcal{F}_{(E)}$.\qed

\begin{lemma}\label{lemmar}
Let $(E)$ be a finite satisfiable set of experimental
equations such that the size of every equation is at most $k$,
and let $\super{G}$ be a CPSO. For every $\eps\geq 0$,
if $\distso(\super{G},\mathcal{F}_{(E)})\leq\eps$
then $\super{G}$ ($k\eps$)-satisfies $(E)$.
\end{lemma}
Our main result is the consequence of Theorems \ref{general},
\ref{couple}, \ref{cnot},
\ref{robustness}, \ref{hadamardrob}, \ref{robtotester},
and the many-qubit generalizations of them.
\begin{theorem}\label{maintheo}
Let $\mathcal{F}$ be one of the following families~:
\begin{itemize}
\item $\mathcal{R}_{\alpha,\theta}\ $ for
$(\alpha,\theta)\in(0,\pi]\times(0,\pi/2]\backslash\{(\pi,\pi/2)\}$
where $\alpha/\pi$ is rational,
\item $\{(\super{H}_\phi,\super{\neg}_{\phi}) : \phi\in[0,2\pi)\}$,
\item $\mathcal{H}\times\{\super{R}_{\pm\alpha}\}\ $ for $\alpha/\pi$
rational,
\item $\{(\super{H}_\phi,\super{\cnot}_{\phi})
: \phi\in[0,2\pi)\}$,
\item $\{(\super{H}_\phi,\super{R}_{s\pi/4},\super{\cnot}_{\phi})
: \phi\in[0,2\pi),s=\pm 1\}$.
\end{itemize}
Then there exists an integer $k\geq 1$ and a real $C>0$ such that,
for all $\eps>0$, $\mathcal{F}$ has an $(\eps,C\eps^{1/k})$-tester
which makes $O(1/\eps^2)$ queries.
Moreover, for every
$0< \eps\leq 1$, $\mathcal{H}$ has an $(\eps/6,4579\sqrt{\eps})$-tester
which makes $O(1/\eps^2)$ queries.
\end{theorem}
Note that each triplet of the last family forms a
universal and fault-tolerant set of quantum gates\cite{bmprv99}.

\section{Acknowledgements}
We would like to thank Jean-Benoit Bost,
St\'ephane Boucheron, Charles Delorme,
St\'ephane Gonnord,
Lucien Hardy, Richard Jozsa, and Vlatko Vedral for several
useful discussions and advice.

This work has been supported by C.E.S.G., Wolfson College Oxford,
Hewlett-Packard, European TMR Research Network ERP-4061PL95-1412,
the Institute for Logic, Language and Computation in Amsterdam,
ESPRIT Working Group RAND2 no. 21726, NSERC, British-French
Bilateral Project ALLIANCE no. 98101, and the
Quantum Information Theory programme of the European Science Foundation.



\appendix
\newcommand{\lemmatwo}{Lemma~\ref{lemma2}}
\section{Appendix: Proof of \lemmatwo{\label{APL2}}}
\textsc{Proof of~\reflemma{lemma2}.}
Let $P$ be the plane defined in $\R^3$ by $\ball{\rho_1}$,
$\ball{\rho_2}$ and $\ball{\rho_3}$. To simplify the discussion, we
suppose w.l.o.g. that $\ball{\zeta^{\pm}_z}$ and
$\ball{\zeta^{\pm}_x}$ are in $P$. Every one-qubit $\rho$
satisfying $\ball{\rho}\in P$ is a linear combination of $\rho_1$,
$\rho_2$ and $\rho_3$. Therefore by linearity of $\super{G}$ we get
that it acts as the identity on $\{\rho : \ball{\rho}\in
P\}^{\otimes n}$. Moreover it is sufficient to show that
$\super{G}$ is the identity on density matrices representing
non-entangled pure states, since they form a basis for all density
matrices.

For this, for every $k$, let $A_k$ be the set of density matrices
representing $k$-qubit non-entangled pure states, and let
$B_k=\{\zeta_x^{\pm},\zeta_z^{\pm}\}^{\otimes n}$.
We will show by induction on $k$ that, for every $0\leq k\leq n$,
the CPSO $\super{G}$ acts as the identity on $A_k\otimes B_{n-k}$.
The case $k=0$ follows by the hypothesis of the lemma.

Suppose the statement is true for some $k$. Fix $\sigma\in A_k$ and
$\tau\in B_{n-k-1}$. For every one-qubit density matrix $\rho$ let
$\tilde{\rho}$ denote the $n$-qubit density matrix
$\sigma\otimes\rho\otimes\tau$.

We now prove that
$G(\tilde{\rho})=\tilde{\rho}$, for every $\rho\in A_1$.
For this, we use the fact that
the density matrix $\Psi^{+}$ representing
the entangled EPR state $(\ket{00}+\ket{11})/\sqrt{2}$,
can be written in terms of
tensor products of the $\zeta$ states:
\begin{eqnarray*}
\Psi^+ & = &
\half(
\zeta_x^+ \ox \zeta_x^+ +
\zeta_x^- \ox \zeta_x^- +
\zeta_z^+ \ox \zeta_z^+ +
\zeta_z^- \ox \zeta_z^-)
-\half(\zeta_y^+ \ox \zeta_y^+ +
       \zeta_y^- \ox \zeta_y^-).
\end{eqnarray*}
This can be generalized for the pure state
$\ket{\mu}=(\ket{\tilde{0}}\ket{\tilde{0}}
+\ket{\tilde{1}}\ket{\tilde{1}})/\sqrt{2}$:
\begin{eqnarray*}
\mu & = &
\half(
\tilde{\zeta}_x^+ \ox \tilde{\zeta}_x^+ +
\tilde{\zeta}_x^- \ox \tilde{\zeta}_x^- +
\tilde{\zeta}_z^+ \ox \tilde{\zeta}_z^+ +
\tilde{\zeta}_z^- \ox \tilde{\zeta}_z^-)
-\half(\tilde{\zeta}_y^+ \ox \tilde{\zeta}_y^+ +
       \tilde{\zeta}_y^- \ox \tilde{\zeta}_y^-).
\end{eqnarray*}
If we apply the CPSO $\super{I}_{2^n} \ox \super{G}$ to the
state $\mu$ we get:
\begin{eqnarray*}
(\super{I}_{2^n} \ox \super{G})(\mu) & = &
\half(
\tilde{\zeta}_x^{+} \ox \tilde{\zeta}_x^{+} +
\tilde{\zeta}_x^{-} \ox \tilde{\zeta}_x^{-} +
\tilde{\zeta}_z^{+} \ox \tilde{\zeta}_z^{+} +
\tilde{\zeta}_z^{-} \ox \tilde{\zeta}_z^{-}
-\tilde{\zeta}_y^{+} \ox\super{G}(\tilde{\zeta}_y^{+})
-\tilde{\zeta}_y^{-} \ox\super{G}(\tilde{\zeta}_y^{-})).
\end{eqnarray*}
If $\ket{\phi}$ and $\ket{\phi'}$ are orthogonal $n$-qubit pure
states, then let
$\Phi^-_{\phi\phi'}=(\ket{\phi}\ket{\phi'}-
\ket{\phi'}\ket{\phi})/\sqrt{2}$.
Since $\Phi^-_{\phi\phi'}$
is orthogonal
to all symmetric $2n$-qubit pure states of the form $\psi \ox \psi$,
by projecting $(\super{I}_{2^n} \ox \super{G})(\mu)$ to
$\Phi^-_{\phi\phi'}$ we obtain:
\begin{eqnarray*}
\bra{\Phi^-_{\phi\phi'}}
(\super{I}_{2^n} \ox \super{G})(\mu)
\ket{\Phi^-_{\phi\phi'}}
& = &
-\half\bra{\Phi^-_{\phi\phi'}}\tilde{\zeta}_y^+
\ox\super{G}(\tilde{\zeta}_y^+)\ket{\Phi^-_{\phi\phi'}}
 -\half\bra{\Phi^-_{\phi\phi'}}\tilde{\zeta}_y^-
\ox \super{G}(\tilde{\zeta}_y^-)\ket{\Phi^-_{\phi\phi'}}.
\end{eqnarray*}
Since $\super{G}$ is a CPSO, the left-hand side of this equality is
non-negative and in the right-hand side both terms are
non-positive. Therefore for every orthogonal $n$-qubit pure states
$\ket{\phi}$ and $\ket{\phi'}$, we get
\begin{equation*}
\begin{array}{rcccl}
\bra{\Phi^-_{\phi\phi'}}\tilde{\zeta}_y^+
\ox\super{G}(\tilde{\zeta_y}^+)\ket{\Phi^-_{\phi\phi'}} & = &
\bra{\Phi^-_{\phi\phi'}}\tilde{\zeta}_y^-
\ox\super{G}(\tilde{\zeta_y}^-)\ket{\Phi^-_{\phi\phi'}} & = & 0.
\end{array}
\end{equation*}

A straightforward calculation then shows that
$\super{G}(\tilde{\zeta_y}^{\pm})=\tilde{\zeta}_y^{\pm}$. Therefore
$\super{G}$ acts as the identity on density matrices
$\tilde{\zeta}_z^{\pm}$, $\tilde{\zeta}_x^+$ and
$\tilde{\zeta}_y^+$, which generate all density matrices, and thus
$\super{G}(\tilde{\rho})=\tilde{\rho}$.
\qed

\newcommand{\theohad}{Theorem~\ref{hadamardrob}}
\section{Appendix: Proof of \theohad\label{APTH}}
\begin{fact}\label{cnxnormso}
Let $\super{G}$ be a superoperator on $\C^{2\times 2}$. Let
$0\leq\eps\leq 1$ be such that
$\normone{\super{G}(\zeta_x^\pm)-\zeta_x^\pm}$,
$\normone{\super{G}(\zeta_y^\pm)-\zeta_y^\pm}$,
$\normone{\super{G}(\zeta_z^\pm)-\zeta_z^\pm}\leq\eps$;
then $\normso{\super{G}-\super{I}_2}\leq 8\eps$.
\end{fact}
\begin{lemma}\label{lemma5}
Let $\ball{u}$ and $\ball{v}$ be two orthonormal vectors in $\R^3$,
and $0\leq\eps\leq 1$ a constant. If $\super{G}$ is a CPSO for one
qubit such that
$\norm{\affine{G}(\pm\ball{u})-\pm\ball{u}}\leq\eps$ and
$\norm{\affine{G}(\pm\ball{v})-\pm\ball{v}}\leq\eps$, then
$\normso{\super{G}-\super{I}_{2}}\leq 241\eps$.
\end{lemma}
\proof
We can suppose w.l.o.g. that $u=\zeta_x^+$ and $v=\zeta_z^+$. Let
$\rho=\super{G}(\zeta_y^+)$, where $\ball{\rho}=(x,y,z)$. {From}
\reflemma{lemma1} it follows that
$\normone{\super{G}(\zeta_z^+)-\rho}
\leq \normone{\zeta_z^+-\zeta_y^+} = \sqrt{2}$.
By the assumption of this lemma we have that
$\normone{\super{G}(\zeta_z^+)-\zeta_z^+}\leq \eps$, and hence
$\normone{\zeta_z^+-\rho}\leq \sqrt{2}+\eps$. The same relation
holds also for the other three
fixed points $\zeta_z^-$, $\zeta_x^+$, and $\zeta_x^-$. As a
result, the three coordinates of $\ball{\rho}$ have to obey the
four inequalities
\begin{equation}
\begin{array}{rcccl}\label{eq:xyzrestriction}
x^2+y^2 + (z\pm 1)^2
\textrm{ and }
(x\pm 1)^2+y^2 + z^2 &
\leq &
(\sqrt{2}+\eps)^2
& \leq &
2+4\eps
\end{array}
\end{equation}

A second set of restrictions on $(x,y,z)$ comes from the complete
positivity of $\super{G}$. Again we use the decomposition of the
EPR state $\Psi^+$, to analyze the two-qubit state:
\begin{eqnarray*}
(\super{I}_2 \ox \super{G})(\Psi^+)
& = &
\half(
\zeta_x^+ \ox \super{G}(\zeta_x^+) +
\zeta_x^- \ox \super{G}(\zeta_x^-))
+\half(
\zeta_z^+ \ox \super{G}(\zeta_z^+) +
\zeta_z^- \ox \super{G}(\zeta_z^-)) \\
&&
-\half({\zeta_y^+ \ox \super{G}(\zeta_y^+)
+\zeta_y^- \ox \super{G}(\zeta_y^-)}).
\end{eqnarray*}
Using the hypothesis, the projection of this state onto the
anti-symmetrical entangled qubit pair $\ket{\Phi^-} =
(\ket{01}-\ket{10})/\sqrt{2}$ yields
\begin{eqnarray*}
\bra{\Phi^-}
(\super{I}_2 \ox \super{G})(\Psi^+)
\ket{\Phi^-}
& \leq & 2\eps
-\half\bra{\Phi^-}\zeta_y^+
\ox\super{G}(\zeta_y^+)\ket{\Phi^-}
-\half\bra{\Phi^-}\zeta_y^-
\ox \super{G}(\zeta_y^-)\ket{\Phi^-}.
\end{eqnarray*}
Since $\super{G}$ is a CPSO,
as in \reflemma{lemma2} we get
$\bra{\Phi^-}\zeta_y^+\ox\rho\ket{\Phi^-} \leq 4\eps$.
A straightforward calculation
shows that this last relation is
equivalent
with a restriction on the $y$ coordinate:
$y \geq 1-16\eps$.

This last inequality implies $y^2\geq 1-32\eps$, which combined
with the restrictions of \refeq{eq:xyzrestriction}, leads to the
conclusion that $\left(x\pm 1\right)^2 \leq 2+4\eps-y^2-z^2  \leq
1+36\eps$, and similarly $\left(z\pm 1\right)^2  \leq 1+36\eps$.
The $x$ and $z$ coordinates of $\ball{\rho}$ satisfy $|x|,|z|
\leq 18\eps$.

These bounds imply
\begin{eqnarray*}
&\begin{array}{rcccl}
\normone{\super{G}(\zeta_y^+)-\zeta_y^+}
& = &
\sqrt{x^2+(y-1)^2+z^2}
& \leq &
\sqrt{904}\eps.
\end{array}&
\end{eqnarray*}
The same result can be proved for $\zeta_y^-$.
Therefore by \reffact{cnxnormso} we can conclude the proof.
\qed

\textsc{Proof of~\reftheo{hadamardrob}.}
Let $\super{G}$ be a CPSO which $\eps$-satisfies the equations.
First we will show there is a point $\ball{\rho}\in\BS$
with $z$-coordinate $0$ whose distance from
$\ball{\super{G}(\ketbra{0}{0})}$ is at most $10\sqrt{\eps}$.
The last two equations imply that
$\normone{\super{G}^2(\ketbra{b}{b})-\ketbra{b}{b}}\leq 3\sqrt{\eps}$,
for $b=0,1$. Therefore
$\normone{\super{G}^2(\ketbra{0}{0})-
\super{G}^2(\ketbra{1}{1})}\geq 2 - 6\sqrt{\eps}$,
and by \reflemma{lemma1}(a) we have
$\normone{\super{G}(\ketbra{0}{0})-
\super{G}(\ketbra{1}{1})}\geq 2 - 6\sqrt{\eps}$.
Thus $\norm{\ball{\super{G}(\ketbra{b}{b})}}\geq 1-6\sqrt{\eps}$,
for $b=0,1$.
Let $\tau=\rho(\half,\alpha)$, where
$\super{G}(\ketbra{0}{0})=\rho(p,\alpha)$.
The first equation implies that
$\norm{\ball{\tau}-\ball{\super{G}(\ketbra{0}{0})}}\leq 2\eps$.
Therefore for $\ball{\rho}=\ball{\tau}/\norm{\ball{\tau}}$
we get $\normone{\super{G}(\ketbra{0}{0})-\rho}\leq 10\sqrt{\eps}$.

The point $\ball{\rho}$ on $\BS$ uniquely defines $\phi\in[0,2\pi)$
such that $\ball{\super{H}_\phi(\ketbra{0}{0})}=\ball{\rho}$. One
can verify that ${\super{H}_{\phi}^{-1}\circ\super{G}}$ acts as the
identity with error at most $19\sqrt{\eps}$ on the four density
matrices ${\ketbra{0}{0}}$, ${\ketbra{1}{1}}$,
${\super{H}_\phi(\ketbra{0}{0})}$,
 and ${\super{H}_\phi(\ketbra{1}{1})}$.
{From} \reflemma{lemma5} we conclude that
$\normso{\super{G}-\super{H}_{\phi}}\leq 4579\sqrt{\eps}$.
\qed

\end{document}